# ICT in Universities of the Western Himalayan Region in India: Status, Performance- An Assessment

Dhirendra Sharma[1] and Vikram Singh[2]

[1]University Institute of Information Technology, Himachal Pradesh University,
Shimla, Himachal Pradesh 171 005, India.

[2]Department of Computer Science and Engg, Ch. Devi Lal University,
Sirsa, Haryna 125 055, India.

**Abstract**

The present paper describes a live project study carried out for the universities located in the western Himalayan region of India in the year 2009. The objective of this study is to undertake the task of assessment regarding initiative, utilization of ICT resources, its performance and impact in these higher educational institutions/universities. In order to answer these, initially basic four- tier framework was prepared. Followed by a questionnaire containing different ICT components 18 different groups like vision, planning, implementation, ICT infrastructure and related activities exhibiting performance. Primary data in the form of feedback on the five point scale, of the questionnaire, was gathered from six universities of the region. A simple statistical analysis was undertaken using weighted mean, to assess the ICT initiative, status and performance of various universities. In the process, a question related to Performance Indicator was identified from each group, whose Coefficient of Correlation was calculated. This study suggests that a progressive vision, planning and initiative regarding academic syllabi, ICT infrastructure, used in training the skilled human resource, is going to have a favourable impact through actual placement, research and play a dominant role at the National and International level.

## 1. Introduction

During the past decade there has been a very rapid development in activities related to Information and Communication Technology (ICT), (Sadagopan 1998), in various Universities and Institutions of higher learning. Such a transformation was triggered by more and more awareness of microwaves and tele-communication. Its pervasiveness has affected almost all fields from micro- to astronomical level via class rooms. ICT in respective fields makes education extremely effective, efficient and engaging due to its power of problem solving and application.

The impetus at the national level was set in motion by the Task Force 'Technology Information Forecasting and Assessment Council' (TIFAC), for Technology Vision 2020. One of the important pilot documents "India 2020, A Vision: for the New Millennium" (1998) prepared by Technology Information Forecasting and Assessment Council (TIFAC), under the leadership of Dr. A.P.J.Kalam, provided a blue print of the `Technology Vision for India'. According to this document ICT can enhance critical thinking, information handling skills, level of perception, problem solving capability and adding value to research in educational institutions. It not only highlighted the importance of higher educational institutions/ universities to take prompt and appropriate initiatives in this direction, but has given further direction for planning ICT strategies in which the role of higher educational institutions in India is going to be very crucial. It emphasized that the Indian human resource has great learning capabilities with a core competence in technology along with the spirit of entrepreneurship and competitiveness. It strongly advocated the development of human resource cadre and the role of higher educational Institutions that will be the foundation of the technological advancement in the country.

This prompted Government of India to take a major ICT initiative to lay down the ICT policy for whole of the country, which is reflected in the planning and implementation by the Ministry of Human Resource Development (MHRD) at various levels particularly through higher educational institutions/ universities of the country. At the national level, it is being implemented vigorously in the form of e-governance and is being coordinated by National Informatics Centers (NIC) throughout the country. University Grants Commission (UGC), All India Council of Technical Education, Department of Science & Technology (DST), and the Indian Council of Agricultural Research (ICAR) are also encouraging higher education Institutions and Universities in this direction. Latest telecommunication policy (Tandon 2002) is quite consistent with the initiatives mentioned above.

In Jan. 2009, the Cabinet Committee on Economic Affairs, Government of India, gave its approval for a new scheme by the name of National Mission on Education through Information and Communication Technology (NMEICT) submitted by the MHRD, Govt. of India, for the 11th Five Year Plan. This mission has been envisaged to leverage the potential of ICT, in providing high quality personalized and interactive knowledge modules for all the learners in higher education institutions in 'any time any where mode'. Further, it has to ensure on-line access to all the ICT resources viz. e- contents, the connectivity and virtual laboratories, to 18,000 colleges and all the universities of the country and thus to bridge the digital divide.

There has been another body, National Association of Software and Services Companies (NASSCOM), in the field of ICT, dealing with the ICT infrastructure, research & development and trade since 1988. According to recent report of NASSCOM (Strategic Review-2009), it has emphasized the availability of skilled and employable talent. In order to achieve this concentrated effort to enhance talent availability and quality are needed from all the concerned sections namely the government academia and the industry. The role of academia is supposed to be very critical in setting up research laboratories. Specific initiatives like faculty development program, upgrading the curriculum, launching internship program and academia- industry collaboration can help to bridge the gap towards the development of talent. NASSCOM is particularly interested in the ICT- business process outsourcing (ICT-BPO) industry in India, which has become a grown economy of the country. Further, this body is a partner with the government of India and various state governments of the country. It has also played a crucial role in the formulation of ICT policies which endeavors to narrow down the digital divide in the country and enables all citizens to enjoy the benefits of ICT.

At the international level, United Nations Educational, Scientific and Cultural Organization UNESCO (2007- 08) has also prepared a document for Asia– Pacific countries, for the implementation of ICT programs through the higher educational institutions. Various ICT strategies of driving the higher education towards excellence in ICT is described in the above document, in support of the core areas of higher education. i.e., learning, teaching, research and training programs.

All these innovative ideas could propagate through the concept of diffusion and adoption. It seems that adoption of the latest information and communication technology is making all the difference particularly in the universities. During the last two decades there has been several attempts to understand information technology diffusion. At the empirical level, the concept of diffusion was reviewed by Fichman in 1992. He has dwelled upon innovation diffusion theory, in particular, how to improve technology assessment, adoption and implementation. A framework was discussed in terms of classical diffusion adopted mainly by individuals and organizations alongwith achieving critical mass beyond which the innovation is universally adopted. A lot of scientific work exists on the diffusion of innovation and its adoption given by Attewell (1992), Rogers (1995), Farquhar & Surry (1994), Anderson et al (1998) and Wilson et al (2000). Attewell (1992) put the work on diffusion in two categories a) Adoption studies and b) Macro diffusion studies. The former is primarily concerned with understanding of innovations and its assimilation during a time of adoption. The latter is concerned with characterizing the rate a pattern of adoption of technology. This type of work was understood in terms of mathematical models of the diffusion process by Mahajan & Peterson (1985) and Mahajan et al. (1990). Innovation diffusion has also been understood in terms of mathematical approaches given by Karmeshu and Pathria (1980), Lal et al (1988), Karmeshu et al (1992) and by Karmeshu and Jain (1995).

In recent years, a very significant work on 'Global Diffusion of the Internet' by Wolcott and Goodman (2003) appeared in Indian context. Walcott and Goodman (2003), presented a vision of new India as a measure of IT power, fully integrated with a global economy. The key to this vision was obviously the internet for enabling the technology based changes. They provided an analytical framework which broadly consisted of dimensions and determinants. Dimensions contained six variables namely organizational infrastructure, geographic dispersion, connectivity infrastructure, pervasiveness, sectoral absorption and sophistication of use. These variables are supposed to capture the state of internet within a country at a given point of time. Each of these variables was judged at five different levels. Determinants represent various factors, to understand the observation of the "state". While discussing the IT action plan, they further elaborated on distinctive features of the internet in India, for a continued dynamic growth, in three steps (a) The government policy (b) The nature of relationship between the government, the state owned telecommunications service providers and private sectors as a critical variable. (c) The policy makers try to strike a balance between the interest of the society and those of individuals. Later, this framework has been applied to understand the initiative, status and performance in Ghana by Foster, et al (2004) in Togo by Bernstien and Goodman (2005) and in Kenya by Ochara et al (2008). Such a comprehensive analysis was carried out in a broader context at the national level for different countries.

In view of the above initiatives from different quarters, most of the universities and higher educational institutions have started adopting the latest technology, focusing on the development of the skilled human resource, as part of its responsibility, in the field of ICT, encompassing all the disciplines, particularly in technical fields, productively and constructively. Most of the universities are undertaking the job of training the technical personnel's in the respective specializations or in the area of its excellence. These programs have been going on for the last almost one decade.

Almost after a decade, one may wonder how efficient the system, particularly the university/higher educational institutions has become. Initiatives to have an assessment of ICT infrastructure, its utilization and overall performance and efficiency, of any higher educational institution, are quite essential, in identifying, planning and achieving the ICT strategies at the respective levels. Essentially, this has formed the basis of motivation to undertake the study of the ICT initiatives, status and its academic performance in various universities in the western Himalayan region of the country and then to analyze, using simple statistical methods, the ICT based academic/ technological status and performance of different universities, and assess in a comparative manner, their vision, planning, initiative, status activities and impact.

The University Grants Commission (UGC), New Delhi, took steps to establish an autonomous institution, National Assessment and Accreditation Council (NAAC), for comprehensive assessment of various universities and to place them according to certain ranking. NAAC (2007) has developed a framework for higher education based on the promotion and sustenance of quality of teaching- learning, research and training programs. Their most significant core value is quest for excellence/ innovations using the latest technological trends and fostering global competence among students. They have devised seven assessment criteria namely, curricular, teaching- learning, research & application, innovative, infrastructure, student support and leadership/governance, aspects, to capture the micro- level quality indicators by using differential weightages. Some other national agencies TOI group 2007, also tried to assess, independently, various agriculture, horticulture and technical institutions. These recommendations and respective gradations are available at the national level. This information may be utilized for comparison with the findings of this paper in the context of ICT.

Recently Mokhtar et al (2007), presented the state of academic computing in Malaysian colleges in which they tired to answer the central questions regarding indicators of assessments and performance of academic computing. They adopted the value chain concept originally proposed by Porter (1985), in connection with some business idea, to describe the relationship between academic ICT activities. This framework consisted of two groups namely primary activities and support activities. Primary activities included use of ICT in learning, teaching, research and training as the core service area and then for enhancing the value in each. It further included remote access to data, faster and more precise data processing, simulation of complex systems, collaboration between research groups. The support activities were linked with primary activities to improve the effectiveness/ efficiency and contain ICT vision, policies and standards, ICT infrastructure, ICT services. Backer and Mohamed (2008) elaborated on the benefits of ICT in teaching. These studies have certain limitation keeping same immediate objectives in mind, in that country. However, the analysis presented by Mokhtar et al and Bakar & Mohamed was academic in nature, carried out for colleges only and not for universities and higher educational institutions of training and research.

We plan to undertake this work for universities of western Himalayan region in India. This region is typically a difficult terrain in which people have to really struggle for livelihood, health and education. It is in this context the government of India is paying special attention to alleviate the level of education using the full potential of ICT, in this region. In view of this, a little different framework for the universities of this region is required. A framework will be developed for assessing the initiative, status and performance of different universities. Initiative will imply vision and planning of the ICT programs in the universities. ICT status would mean complete ICT infrastructure including local area network facility, internet network security, mobile computing access, system application software, website and information system, teaching display technologies, ICT technical staff, ICT budget allocation, e- library, e- placement/ alumni portal. ICT performance will be related with various activities based on ICT infrastructure. It would include teaching; learning and ICT based research at the conventional level and at the advanced level in the form of ICT training programs. This performance is going to have an impact of the ICT programs, functional in the universities, in the form of actual placements of the outgoing students, research publications and actual problem solving.  This outline of the framework will be discussed explicitly in the next section.

 The objective of this study is   to undertake the task of assessment regarding initiative, utilization of ICT resources, its performance and impact in these higher educational institutions/ universities. In order to achieve this, an overall status report is prepared using the framework of this paper. A questionnaire was prepared accordingly, related to various ingredients of ICT, initiative, status and performance. Then a survey was conducted in different  universities located in the western Himalayan region, as per the questionnaire. The analysis and comparison of the initiative, status and performance through performance indicators in respective universities has been carried out.  The corresponding benchmarking is expected to identify the future ICT strategies in order to improve their status as well as the product in the form of professionally skilled human resource in the field of ICT. The reference period of the study has been the year 2009.

The paper is organized in six sections. After the introduction in section 1, the basic framework will be presented in section 2. The methodology, preparation of questionnaire, a survey and analytic procedure will be given in section 3.  Section 4 will deal with the results obtained from the analysis of the primary data. Summary and conclusions are given in Section 5. Future strategies are discussed in the final Section.

## 2. Basic Framework

Before discussing the methodology it seems proper to have a framework on the basis of which the study of ICT initiative, status, performance and impact in the field of ICT in different universities and higher technical institution, is carried out.
The important building blocks are supposed to be Performance Indicators (PI). One group of authors (Riley and Nuttall 1994) suggests that Performance Indicators must be something quantitative, whereas the other group takes a wider view in favour of qualitative and descriptive statement for PI. The latter view is adopted by International Standards Organisation (1998). Both these measures, quantitative as well as qualitative, will be used here that will allow us   to have a complete view of richness and diversity of the ICT based academic/ technological performance and the related activities. With these it should be possible to assess and judge the effectiveness, efficiency and impact of an institution in terms of ICT performance to have the possible projection to finally decide about the future strategies.

In deciding the basic framework, the question is what are the  ICT ingredients/ components and Performance Indicators for assessing the ICT initiative, status and their performance at different levels.  The basic ingredients of ICT are clearly classified among four tiers, containing 117 questions divided in 18 groups, A-R. These groups, in their respective tier, depict different academic components of ICT and the related activities. Schematic representation of four tier framework is given in Fig. 1.

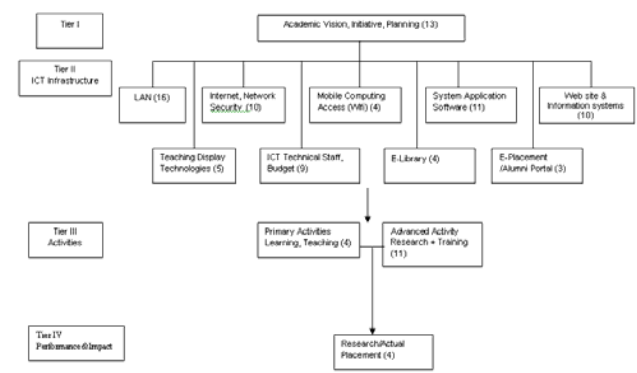

Fig. 1: Four Tier framework. Each box contains the main ICT component along with the number of questions in it

The basis of formulating this framework has been the recommendations contained in the documents available at the national and international level, as mentioned in the introduction. This framework is supposed to be most suited to the university and higher educational institutions which are mainly the centre of research where knowledge is created. In each group one of the questions, the last one, is related to the performance. All these performance related questions, are placed in another group, S. These groups, alongwith the Performance Indicator in each, are presented in Table I:

Table I: Various Groups of ICT Components in an Institution.

| S.No | Group | ICT details at University level | Performance Indicators | No. of Questions | Tier in Framework |
|---|---|---|---|---|---|
| 1. | A | ICT Vision | Implementation | 03 | TIER-I |
| 2. | B | ICT Initiatives | Execution | 04 | |
| 3. | C | ICT Planning | Achievement | 06 | |
| 4. | D | ICT Infrastructure | Utilization | 13 | TIER-II |
| 5. | E | Local Area Networks Facility | Users | 16 | |
| 6. | F | Internet, Network Security | Speed | 10 | |
| 7. | G | Mobile Computing access | Access 24x7 basis | 04 | |
| 8. | H | System Applications S/W | Overall user's Satisfaction | 11 | |
| 9. | I | Web site and information systems | Online communication | 09 | |
| 10. | J | Teaching Display Technologies | Availability | 05 | |
| 11. | K | ICT Technical Staff | Trouble Shooting | 05 | |
| 12. | L | ICT Budget allocation | Utilization | 04 | |
| 13. | M | E- Library | Availability | 04 | |
| 14. | N | E-Placement/Alumni Portal | Alumni Portal | 03 | |
| 15. | O | ICT based Teaching and learning | Effective use of ICT | 04 | Tier III |
| 16. | P | ICT Based research | Satisfaction level | 07 | |
| 17. | Q | ICT Training programmes | No. of programmes organized | 04 | |
| 18. | R | Impact of ICT: Research & Actual Placement | Placement /Research | 04 | TIER IV |
| 19. | S | Performance | Success rate | 18 | |

## 3. Methodology

The universities located in the western Himalayan region imply those spreading over the state of Himachal Pradesh and Jammu & Kashmir. There are mainly 7 universities located in the western Himalayan region of India, out of which 6 universities have been selected purposely. Five universities belong to the state or the central government, and one is a private university but deemed university approved by the University Grants Commission under the aegis of Ministry of Human Resource Development, Government of India. These are Himachal Pradesh University (Shimla), Y.S. Parmar Horticulture University (Solan), C.S.K. Agriculture University (Palampur), National Institute of Technology (Hamirpur), Jammu University (Jammu) belonging to the government and J.P. University (Solan) as deemed university.

In order to have the information regarding ICT initiative, status and performance in the universities of this region, the questionnaire was given to the concerned head of the department of ICT/ IT/ Computer Science in the institution/ universities personally. The concerned head was requested to provide information/ feed back as per the questionnaire. Thus the information was gathered personally by the authors.

In the structured questionnaire, the feedback to various queries were on five point scale arranged in a particular order that revealed the natural flow from the lowest level to the highest level, in increasing order of sophistication. That is why it was thought reasonable to give a weightage of 1 to 5 respectively to each level in increasing order.

Simple standard statistical tools were used to analyse the data groupwise to find the Weighted Mean, Standard Deviation and Coefficient of Variation (CV) keeping in mind the weight of respective levels. The CV indicates the variation around its weighted mean in the series, the lesser is the CV, more consistent and stable is the series. However, the weighted mean was found to be a better measure as compared to CV, to understand the trend in a particular group. The use of median was another alternative but in view of the fact that the data was not having extreme variation and due to the limitation of the median method, weighted mean was logically preferred. Another statistical quantity 'Coefficient of Correlation' among various groups was also calculated which has also been used to ascertain the proper groupings in the questionnaire.

A simple tabular analysis was carried out to find out the results. In view of the fact that queries were replied on five point scale with the respective weight from one to five, firstly weighted mean of each group was calculated followed by overall mean in the respective tier. The weighted mean was found by picking up the proper value of the response in a particular level on the five point scale (say L1 to L5). Then dividing it by the total value of the levels, i.e. sum of 1 to 5, multiplied by the number of questions in the group. This is repeated for all the groups for six universities of the region. These are presented in Table II. The Pearson's Coefficient of Correlation of each group with the Performance Indicator group is given in the last column of the Table II.

Table II: Weighted Mean and Pearson's Correlation Coefficients of various groups with Group R

| Tier | Groups | J.P.U Solan | NIT Hamirpur | CSK.U Palampur | J.U. Jammu | Y.S.P.U. Nauni | H.P.U, Shimla | Correlation |
|---|---|---|---|---|---|---|---|---|
| Tier I | Group A | 0.3 | 0.33 | 0.17 | 0.3 | 0.27 | 0.33 | 0.60 |
| | Group B | 0.27 | 0.27 | 0.2 | 0.22 | 0.18 | 0.28 | 0.70 |
| | Group C | 0.29 | 0.28 | 0.21 | 0.21 | 0.24 | 0.32 | 0.48 |
| Weighted. Mean | | 0.29 | 0.30 | 0.19 | 0.24 | 0.23 | 0.31 | |
| Tier II | Group D | 0.33 | 0.33 | 0.28 | 0.22 | 0.22 | 0.3 | 0.70 |
| | Group E | 0.32 | 0.33 | 0.21 | 0.23 | 0.24 | 0.2 | 0.93 |
| | Group F | 0.27 | 0.28 | 0.23 | 0.26 | 0.19 | 0.28 | 0.64 |
| | Group G | 0.2 | 0.29 | 0.23 | 0.24 | 0.17 | 0.17 | 0.60 |
| | Group H | 0.07 | 0.33 | 0.07 | 0.27 | 0.16 | 0.07 | 0.49 |
| | Group I | 0.24 | 0.24 | 0.16 | 0.18 | 0.19 | 0.19 | 0.93 |
| | Group J | 0.15 | 0.27 | 0.15 | 0.14 | 0.07 | 0.12 | 0.80 |
| | Group K | 0.24 | 0.33 | 0.18 | 0.16 | 0.13 | 0.18 | 0.92 |
| | Group L | 0.33 | 0.27 | 0.2 | 0.2 | 0.09 | 0.11 | 0.77 |
| | Group M | 0.23 | 0.33 | 0.23 | 0.2 | 0.23 | 0.17 | 0.71 |
| | Group N | 0.31 | 0.27 | 0.13 | 0.17 | 0.23 | 0.17 | 0.79 |
| Weighted Mean | | 0.24 | 0.30 | 0.10 | 0.20 | 0.17 | 0.18 | |
| Tier III | Group O | 0.27 | 0.27 | 0.25 | 0.2 | 0.26 | 0.15 | 0.38 |
| | Group P | 0.2 | 0.27 | 0.07 | 0.18 | 0.07 | 0.23 | 0.95 |
| | Group Q | 0.2 | 0.2 | 0.09 | 0.12 | 0.07 | 0.08 | 0.91 |
| Weighted Mean | | 0.22 | 0.25 | 0.14 | 0.17 | 0.13 | 0.15 | |
| Tier IV | Group R | 0.3 | 0.32 | 0.17 | 0.23 | 0.15 | 0.22 | 0.63 |
| | Group S | 0.28 | 0.32 | 0.17 | 0.21 | 0.18 | 0.2 | |

## 4. Results and Discussions:

The weighted mean for Tier I, vision, initiative and planning, for different universities has been presented in Fig.2, which depicts a comparative view among the universities.

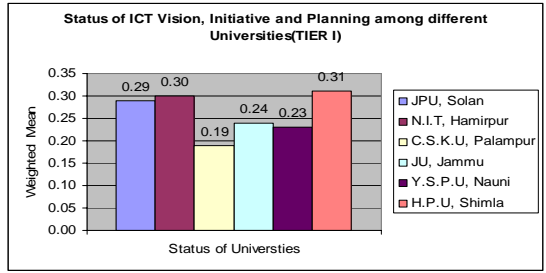

Fig.2 Tier I.

This diagram shows that there is not much difference corresponding to the vision and planning, particularly in the first three universities i.e. JP University, NIT and H.P.University. At the level of vision, initiative and planning all the universities thought big with a minor difference. Standard deviation and the coefficient of variation were also calculated. A range of coefficient of variation (%) was found to be 14.54%, 15.13%, 18.04%, 24.05, 32.66% and 39.44% respectively for H.P.University, J.P.University, NIT, Hamirpur, Nauni University, Jammu University and Palampur University. It reveals that first three universities are having more consistency in ICT vision, initiative and planning than the later three universities. Because for the same vision, the ICT initiatives and planning were at different levels. Further, the level of vision also differed from one university to another.

The overall weighted mean of all the different groups related to Tier II which mainly deals with overall ICT infrastructure among different universities in the region, has been presented in Fig. 3.

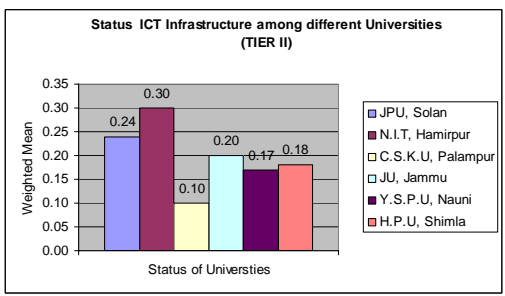

Fig. 3. Tier II

From this diagram, it may be seen that NIT Hamirpur is having the largest ICT infrastructure followed by J.P. University, Jammu University, H.P.University, Nauni University and Palampur University. This difference may be mainly attributed to advanced ICT facilities available in NIT, Hamirpur like better Local Area Network (LAN) with internet facilities, video-conferencing facility available with each faculty, Mobile (Wi-Fi) computing facility, IP telephony and assured access to all the digital tools/ resources on the campus. These facilities are also available in the hostels and faculty accommodation as well, along with an effective web site and information system. In this respect NIT Hamirpur has an edge over other universities. Other contributing factors are the effectiveness of ICT training progammes, organized for the faculty, students and professionals as the drive for human resource development programs. Finally, the placement of professionals in reputed establishments adds much more weightage in this direction.

This figure further shows that the difference between Jammu University and Himachal Pradesh University is very little. It is particularly because of the fact that Jammu University is having higher internet bandwidth, better mobile computing facilities alongwith e-library as compared to those available in Himachal Pradesh University.

Tier III, is concerned with the main ICT activities which can be further divided into primary activities of teaching and learning and advanced activities like sophisticated training & research using ICT for developing professional skills. These are displayed in fig 4.

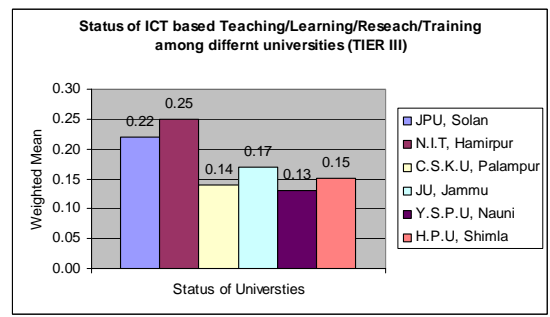

Fig.4 Tier III.

This category deals with various academic activities. Teaching and learning using ICT based tools is one of the effective approaches of procuring knowledge. Researches using e-journals and the material available on the internet have become the lifeline of any good research activity. These ICT tools have provided the academic community a more versatile powered instrument. Further, awareness and training programs at the advanced level to produce professionally skilled technological human resource have become more and more effective using these ICT tools. In this respect, NIT Hamirpur has established superiority over others in this category also, followed by J.P.University, Jammu University and H.P.University etc.

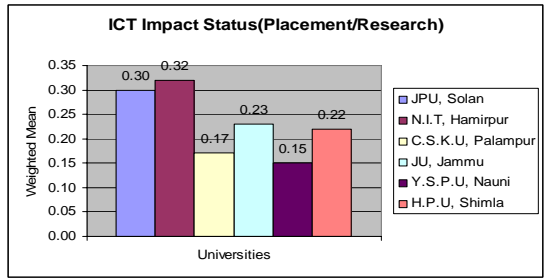

Fig. 5. Tier IV

Tier IV of the framework exhibits the impact of universities at the local (societal) level, at the national and international level as well. The impact is reflected mainly through two components. One is outstanding research supported with patents for the fruitful activities and for the benefit of mankind. Another is the placement of outgoing students trained in professional courses, to reputed universities, institutes, industries and other organizations. The impact factor has been given in Fig.5, which reveals that NIT Hamirpur is again having an edge over other universities in the western Himalayan region of India, followed by J.P.University, Jammu University and Himachal Pradesh University. The main reason for this variation is the value addition through research, particularly supported by the advanced ICT infrastructure. Further, large number of good research publications per year, along with successful collaborative research work also becomes an important factor for the superiority of one university over the other. In addition, actual placement contributes the most to Tier IV in the form of linkages of the professionals with industries.

Towards the end, it was thought reasonable to have Pearson's Coefficient of Correlation (CC) of different groups with the group S depicting the Performance Indicators. These Correlation Coefficients have been given in Table II and displayed in Fig. 6.

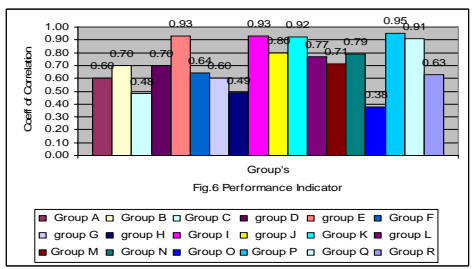

Fig. 6. Correlation with performance Indicators.

On its perusal, the CCs are found to vary between 0.38 and 0.95. Its value, >90, was found for five groups. These groups are E (LAN), I (Website and Information System), K (ICT Staff), P (ICT based Research) and Q (ICT training programs). It speaks for the obvious that the advanced internet facility used in research and training programs has a better correlation with the performance of an institution. This is found quite consistent with the analysis presented above. The next five groups having CCs between 0.70- 0.89, are found to be due to D (ICT Infrastructure), J (Teaching Display Technologies), L (ICT Budget Allocation), M (e-Library) and N (Alumni portal and placement). There is another set of five groups having CCs between 0.6-0.69, which is found just acceptable. These are related with ICT vision, initiative, internet & network security, mobile computing facility and impact of ICT. The remaining three groups having correlations below 0.60 are not acceptable due to inadequate planning in ICT based teaching programs.

## 5. Summary and Conclusion

In this study we have presented an assessment from the primary data gathered as feedback to a questionnaire divided into 18 different relevant groups in respect of ICT activities starting from vision, initiative, planning, implementation, ICT infrastructure, to the performance/ success of various ICT initiatives in six universities in the western Himalayan region. Performance Indicator of each of the groups was identified that was utilized to gauge the success of ICT initiatives at different stages of the framework.

The vision and planning covered in Tier I are not enough if these are not supported with the proper initiative by a visionary at the highest level in the university system with full support from the government along with a small but dynamic team involved with it. It may be emphasized that a good academic curriculum at the teaching level and better faculty are the key factors which are most essential to impart the latest in the field of ICT.

It further requires equally good infrastructure (Tier II) hardware, software and fast access to the internet in the ICT laboratory.

After all the universities are meant (Tier III) to create knowledge and the professionally skilled dedicated manpower trained not at the local level but at the international level. Professional spirit with dedication of a person is further gauged with how much busy his academic schedule remains for understanding and for problem solving.

The better performance always results in sound impact (Tier IV) at the national and international level in terms of outstanding research publications and actual placements.

It may be concluded from the analysis of all the components of the four- tier framework that NIT Hamirpur has a clear- cut superiority over the other universities in this region, due to the sound vision, excellent ICT infrastructure and efficient ICT based activities and fast adoption of innovation in the field of ICT, whose impact became known at the national and international level. The other Universities should follow the live example of NIT Hamirpur.

Finally, it is the performance in each of the groups that really matters. The correlation of Performance Indicators (group S) with all the other eighteen groups echoes the same. These correlation coefficients further provided the strength to the conceptualization of groups presented in the questionnaire used in this work.

Our findings about the overall ICT status is on the basis of performance and impact do match with the reported ranking given to universities by NAAC and other agencies. This finding is quite interesting which proclaims that overall progress shown by any of the universities is mainly due to the respective progress through initiative, planning, infrastructure and performance in the field of ICT. One could not think of scoring better ranking at the national or international level without the proper development in ICT which has become synonym with the progress of the university.

## 6. Future Strategies

The pattern revealed in this paper in respect of initiative, status and performance of different universities in the targeted area of the country points towards future strategies as follows:

Although ICT initiatives in different universities are quite encouraging, its extension must be envisaged in all higher educational institutions of the country. It must include the advanced training programs to prepare professionally skilled manpower using ICT for enhancing the core competence of scientists in the field of ICT and making these programs more meaningful.

These skilled professionals should function as nucleus to further extend such programs to create more skilled human resource.

They must plan with determination, of removing the sufferings of humanity using the earned knowledge to fulfill the objectives of `Transforming the Society' as envisaged in the Kothari Commission report of 1966 and the Technology Vision 2020 document prepared by TIFAC for India

Different courses in the form of e-content inclusive of video, in Indian universities are made available as freeware similar to those of MIT open courseware, as envisioned by NMEICT. It will have an effective long term sustainable impact on society by way of advance ICT enabled education and empowerment of people in the western Himalayan region of the country.

### Acknowledgments

Authors express their sincere thanks to Dr. S.P. Saraswat, Agro-Economic centre, H.P.University, Shimla, for very fruitful discussion during data analysis.

**Dhirendra Sharma** is pursuing his Ph.D from the Department of Computer Science and Engineering, Ch. Devi Lal University, Sirsa, Haryana, India. He has obtained his MBA from Maastricht School of Management, Masticht, Netherlands, M.S from BITS, Palani, India and Masters in Physics from Himachal Pradesh University, Shimla, India. His areas of interest are ICT in Educational Institutes, computer networking (wired and wireless), Sensor Networks, and Open source web content management. He played a very important role in the Design and Implementation of Campus Wide Optical Fibre Network backbone of Himachal Pradesh University, Shimla. He is having more than 10 years of teaching experience in addition to his 5 years in IT Industry.

**Dr. Vikram Singh** is Ph.D in Computer Science from Kurukshetra University, Kurukshetra, India. Presently he is working as Professor and Head in the Department of Computer Science & Engg and Dean Faulty of Engineering, Ch. Devi Lal University, Sirsa – 125055.Haryana, (India), since 2004 onwards. Before this he was working with Kurukshetra University, Kurukshetra. His areas of research are Computer Networks, E-Governance, Simulation tools, etc. He is having more than 17 years of teaching/research experience and has written two books and having more than 30 publications in international and national journals/conference proceedings.